\documentstyle[prl,aps,twocolumn,psfig,floats]{revtex}

\begin{document}

\title{Constructing the phase diagram of finite neutral nuclear matter}

\author{
J. B. Elliott$^{ 1 }$, 
L. G. Moretto$^{ 1 }$, 
L. Phair$^{ 1 }$, 
G. J. Wozniak$^{ 1 }$,
S. Albergo$^{ 2 }$, 
F. Bieser$^{ 1 }$, 
F. P. Brady$^{ 3 }$, 
Z. Caccia$^{ 2 }$,\\
D. A. Cebra$^{ 3 }$,
A. D. Chacon$^{ 4 }$, 
J. L. Chance$^{ 3 }$, 
Y. Choi$^{ 5 }$, 
S. Costa$^{ 2 }$,\
M. L. Gilkes$^{ 5 }$, 
J. A. Hauger$^{ 5 }$, 
A. S. Hirsch$^{ 5 }$,\\ 
E. L. Hjort$^{ 5 }$, 
A. Insolia$^{ 2 }$, 
M. Justice$^{ 6 }$, 
D. Keane$^{ 6 }$, 
J. C. Kintner$^{ 3 }$,
V. Lindenstruth$^{ 7 }$, 
M. A. Lisa$^{ 1 }$, 
H. S. Matis$^{ 1 }$,\\
M. McMahan$^{ 1 }$,
C. McParland$^{ 1 }$,
W. F. J. M\"{u}ller$^{ 7 }$, 
D. L. Olson$^{ 1 }$,
M. D. Partlan$^{ 3 }$, 
N. T. Porile$^{ 5 }$, 
R. Potenza$^{ 2 }$, 
G. Rai$^{ 1 }$,\\
J. Rasmussen$^{ 1 }$, 
H. G. Ritter$^{ 1 }$, 
J. Romanski$^{ 2 }$, 
J. L. Romero$^{ 3 }$,
G. V. Russo$^{ 2 }$, 
H. Sann$^{ 7 }$, 
R. P. Scharenberg$^{ 5 }$, 
A. Scott$^{ 6 }$,\\ 
Y. Shao$^{ 6 }$,
B. K. Srivastava$^{ 5 }$,
T. J. M. Symons$^{ 1 }$,
M. Tincknell$^{ 5 }$, 
C. Tuv\'{e}$^{ 2 }$, 
S. Wang$^{ 6 }$, 
P. Warren$^{ 5 }$, 
H. H. Wieman$^{ 1 }$,\\ 
T. Wienold$^{ 1 }$, and 
K. Wolf$^{ 4 }$
}

\address{
$^1$Nuclear Science Division, Lawrence Berkeley National Laboratory,
Berkeley, CA 94720\\
$^2$Universit\'{a} di Catania and Istituto Nazionale di Fisica
Nucleare-Sezione di Catania, 95129 Catania, Italy\\ 
$^3$University of California, Davis, CA 95616\\ 
$^4$Texas A\&M University, College Station, TX  77843\\
$^5$Purdue University, West Lafayette, IN 47907\\
$^6$Kent State University, Kent, OH 44242\\
$^7$GSI, D-64220 Darmstadt, Germany}

\date{\today}

\maketitle

\begin{abstract}
The fragment yields from the multifragmentation of gold, lanthanum and
krypton nuclei obtained by the EOS Collaboration are examined in terms
of Fisher's droplet formalism modified to account for Coulomb
energy.\ \ The critical exponents $\sigma$ and $\tau$ and the surface
energy coefficient $c_0$ are obtained.\ \ Estimates are made of the
pressure-temperature and temperature-density coexistence curve of
finite neutral nuclear matter as well as the location of the critical point.
\end{abstract}


\narrowtext

\section{Introduction}
\label{Introduction}

In past attempts to investigate the relationship between nuclear
multifragmentation and a liquid to vapor phase transition
\cite{finn-82,siemens-83,gilkes-94,pochodzalla-95,campi-97,moretto-97,bonasera-00,dagostino-00,elliott-00.1,elliott-02,bauer-02,srivastava-02}
various studies have sought to determine one or more critical
exponents
\cite{finn-82,gilkes-94,elliott-00.1,elliott-02,bauer-02,dagostino-99},
other studies have examined caloric curves \cite{pochodzalla-95}, and
still others have reported the observation of negative heat capacities
\cite{dagostino-00}.\ \ These studies suffer from the lack of
knowledge of the system's location in pressure-density-temperature
$(p, \rho, T)$ space.\ \ For example, interpretations of caloric
curves and negative heat capacities depend on assumptions of either
constant pressure or constant density \cite{moretto-96,elliott-00.3}.\
\ In the case of determining critical exponents, it was assumed that
the fragmenting system is at coexistence and the dominant factor in
fragment production was the surface energy.\ \ The analysis presented
below makes no assumptions about the location of the system in $(p,
\rho, T)$ space and allows for other energetic considerations with
regards to fragment production.

In this paper the analysis technique recently used on
multifragmentation data collected by the ISiS Collaboration
\cite{elliott-02} is applied to the data sets for the
multifragmentation of gold, lanthanum and krypton nuclei collected by
the EOS Collaboration.\ \ All three EOS experimental data sets are
shown to contain the signature of a liquid to vapor phase transition
manifested by the scaling behavior predicted by Fisher's droplet
formalism, and the liquid-vapor coexistence line is determined over a
large temperature interval extending up to and including the critical
point.\ \ The critical exponents $\tau$ and $\sigma$, as well as the
critical temperature $T_c$, the surface energy coefficient $c_0$ and
the compressibility factor $C_F$ are directly extracted.\ \ From the
behavior of the fragment yields the $p$-$T$ and $T$-$\rho$ coexistence
curves are determined and the critical pressure $p_c$ and critical
density ${\rho}_c$ are estimated.

\subsection{Overview}
\label{overview}

The paper is organized as follows: section \ref{EOS-experiment}
reviews the EOS data sets; section \ref{Fisher's-droplet-formalism}
reviews Fisher's droplet formalism and its connection to nuclear
evaporation; section \ref{fitting-the-data} discusses the details of
the data analysis; section \ref{Results} reports the results of the
data analysis; section \ref{phase-diagram} shows the physical
implications of these results; and finally, in section
\ref{Conclusion} a brief discussion of the results is made.\ \ In
order to demonstrate the efficacy of the analysis performed on the EOS
data sets, an appendix shows the results of this analysis performed on
percolation cluster distributions.

\subsection{EOS data sets}
\label{EOS-experiment}

The EOS Collaboration has collected data for the reverse kinematics
reactions $1.0$ AGeV Au$+$C, $1.0$ AGeV La$+$C and $1.0$ AGeV Kr$+$C
\cite{hauger-98,hauger-00}.\ \ There were ${\sim}25,000$, ${\sim}22,000$
and ${\sim}36,000$ fully reconstructed events recorded for the Au$+$C,
La$+$C and Kr$+$C reactions, respectively.\ \ The term ``fully
reconstructed'' means that the total measured charge in each event was
within three units of the charge of the projectile.

        \begin{figure}
        \centerline{\psfig{file=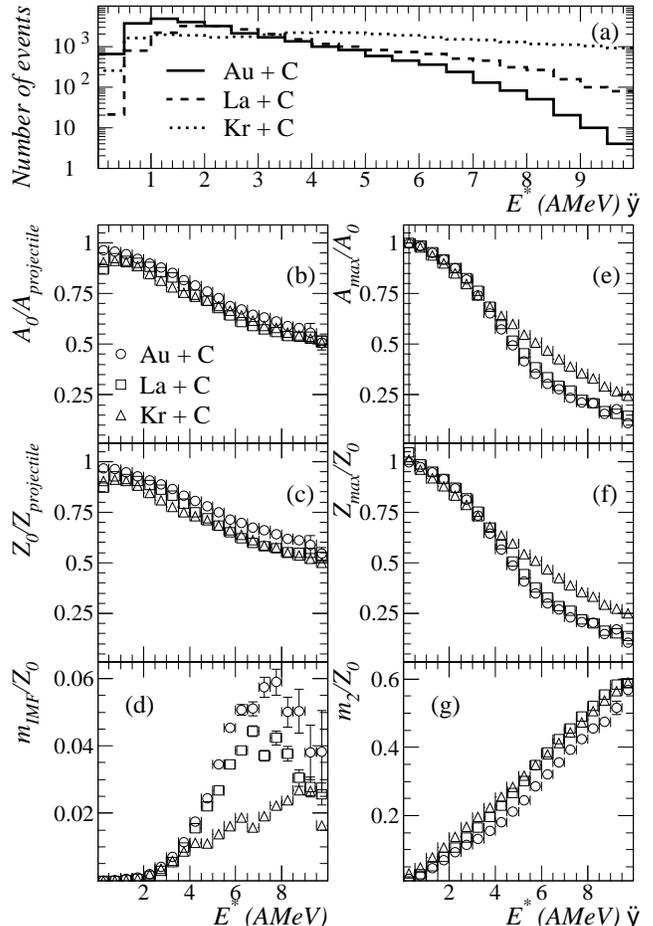,width=8.3cm,angle=0}}
        \caption{(a) The distribution of events as a function of
	excitation energy.\ \ (b) The nucleon number of the
	fragmenting system $A_0$ normalized to the nucleon number of
	the projectile $A_{\text{projectile}}$.\ \ (c) The charge of
	the fragmenting system $Z_0$ normalized to the charge of the
	projectile $Z_{\text{projectile}}$.\ \ (d) The number of
	intermediate mass fragments ($4 \le Z \le Z_0 / 4$)
	$m_{\text{IMF}}$ normalized to the charge of the fragmenting
	system.\ \ (e) The nucleon number of the largest fragment
	$A_{max}$ normalized to $A_0$.\ \ (f) The charge of the
	largest fragment $Z_{max}$ normalized to $Z_0$.\ \ (g) The
	multiplicity of fragments ($1 \le Z \le Z_{max}$) resulting
	from the fragmentation of the system $m_2$ normalized to
	$Z_0$.}
        \label{systematics}
        \end{figure}

For every event, the charge and mass of the projectile remnant
($Z_{0}$, $A_{0}$) were determined by subtracting the charge and mass
of the particles knocked out of the projectile from the charge and
mass of the projectile \cite{hauger-98,hauger-00}.

The thermal component of the excitation energy per nucleon of the
remnant $E^*$ was determined as follows.\ \ First, the total
excitation energy per nucleon $E^{*}_{\text{total}}$ was reconstructed
based on an energy balance between the initial stage of the excited
remnant and the final stage of the noninteracting fragments.\ \ The
prescription \cite{cussol-93} for calculating $E^{*}_{\text{total}}$
is then
	\begin{equation} 
	E^{*}_{\text{total}} = \left[ \sum \left( KE_i + Q_i \right) + \frac{3}{2}nT \right] / A_0 
	\label{energy-balance}
	\end{equation}
where $n$ is the multiplicity of neutrons produced via fragmentation,
$KE_i$ is the kinetic energy of the $i^{\text{th}}$ fragment in the
reference frame of the remnant and $Q_i$ is the removal energy and $T$
is the temperature of a Fermi gas.\ \ For further details see
reference \cite{hauger-98}.

The thermal component of the excitation energy per nucleon of the
remnant $E^*$ was then determined by subtracting the expansion energy
$E_{\text{X}}$ from $E^{*}_{\text{total}}$, where the quantity
$E_{\text{X}}$ is given by
	\begin{equation}
	E_{\text{X}} = \left( \sum KE - E_{\text{th}} - E_{\text{Coulomb}} \right) / A_0
	\label{expansion-energy}
	\end{equation}
with $\sum KE$ the total kinetic energy; $E_{\text{th}}$ the sum of
the translational thermal contribution to the fragment spectra; and
$E_{\text{Coulomb}}$ the Coulomb contribution.

The translational energy is given by
	\begin{equation}
	E_{\text{th}} = \frac{3}{2} T_{\text{isotope}} (m_2+n-1)
	\label{thermal-component}
	\end{equation}
where $m_2$ is the multiplicity of fragments and $T_{\text{isotope}}$
is the temperature calculated from the isotopic yields
\cite{hauger-98,hauger-00}.\ \ This form follows that outlined in
reference \cite{bondorf-95}.

The Coulomb contribution is given by
	\begin{equation}
	E_{\text{Coulomb}} = \frac{3}{5} e^2 \left[ 
	\frac{Z_{0}^{2}}{R_{r}} - \sum_{Z} \frac{N_Z Z^2}{(V_r / V_0)^{1/3}R_Z}
	\right]
	\label{coulomb-component}
	\end{equation}
where $R_r$ is the radius of the excited remnant, $N_Z$ is the number
of fragments with charge $Z$, $R_Z$ is the radius (at normal density)
of a fragment with charge $Z$.\ \ The volumes (and radii) were $V_0$
the volume of the remnant at normal density and $V_r =
V_{\text{projectile}} (T_{\text{Fermi gas}}/T_{\text{isotope}})^{3/2}$
the volume of the excited remnant that isentropically expands from the
normal volume of the projectile $V_{\text{projectile}}$
\cite{hauger-98}; $R_r$ is then determined from $V_r$ assuming a
spherical volume.\ \ This form of $E_{\text{Coulomb}}$ follows
reference \cite{bondorf-95} and takes into account the changing volume
of the excited remnant as a function of excitation energy.\ \ Previous
estimates of $E_{\text{Coulomb}}$ did not account for the changing
volume of the fragmenting remnant \cite{hauger-98,hauger-00}.\ \ This
difference leads to a few AMeV difference in $E^*$ in the most violent
collisions.

For the analysis in this paper, the data for each system was binned in
terms of $E^*$ in units of $0.5$ AMeV; i.e. $20$ bins covered the
excitation energy range $0 \text{AMeV} \le E^* \le 10 \text{AMeV}$.\
\ Figure~\ref{systematics} shows some of the systematics of the EOS
data binned in this manner.\ \ These results are consistent with other
EOS publications \cite{srivastava-02,hauger-98,hauger-00,srivastava-01}.

The systematics of the EOS data sets shown in Fig.~\ref{systematics}
demonstrates the similarity in behavior exhibited by the data sets
when their differing sizes are taken into account by normalizing the
quantity in question to the projectile charge $Z_{\text{projectile}}$
or the charge of the fragmenting system $Z_0$.\ \ The exception is
seen in Fig.~\ref{systematics}d where only below $E^* \sim 4$ AMeV do
all three systems behave similarly.\ \ Above $E^* \sim 4$ AMeV, the
size of the fragmenting systems dominate.\ \ This is reflected in the
ordering of $m_{\text{IMF}}/Z_0$; from lowest to highest: krypton,
lanthanum and gold.

\section{Analysis}
\label{Analysis}

As with several other analyses
\cite{finn-82,gilkes-94,bonasera-00,dagostino-00,elliott-00.1,elliott-02,bauer-02,hirsch-84,goodman-84,mahi-88,campi-89},
the basis of the present effort lies in an examination of the fragment
yield distribution in the context of Fisher's droplet formalism
\cite{fisher-67,fisher-69,stauffer_kiang-70,kiang-70,stauffer_kiang-77}.\
\ Thus, a brief review of Fisher's formalism is given in the following
section, together with a justification for its applicability to
nuclear decay rates.

\subsection{Fisher's droplet formalism}
\label{Fisher's-droplet-formalism}

Fisher's droplet formalism and its forerunners
\cite{mayer-40,frenkel-46} are based on an equilibrium description of
physical clusters or droplets that condense in a low density vapor.\ \
While Fisher's formalism has long been applied to nuclear
multifragmentation yields
\cite{finn-82,gilkes-94,bonasera-00,dagostino-00,elliott-00.1,elliott-02,bauer-02,hirsch-84,goodman-84,mahi-88,campi-89},
the question arises as to the validity of a picture of clusters in
equilibrium within a low density vapor to experiments in which excited
nuclei undergo multifragmentation in vacuum.\ \ Specifically: ``In
which sense is there an equilibrium between liquid and vapor in the
free (vacuum) decay of a (multifragmenting) hot intermediate
(nucleus)?''\ \ Or more to the point: ``Where is the vapor?''

If one assumes, as in a compound nucleus reaction, that the initial
collision entity relaxes {\it quickly} to a hot thermalized blob,
which proceeds {\it slowly} to emit particles stochastically, the
answers to this question is simple.\ \ The hot blob is the liquid
which is evaporating in free space according to standard evaporation
theories.\ \ To establish coexistence, the vapor need not be present.\
\ All that is necessary is to appreciate that:
	\begin{enumerate}
		\item in first order phase transitions the interaction
		      between the two phases is unnecessary; 
		\item the rate of evaporation defines uniquely the
		      vapor phase, even when the vapor phase is
		      absent.
	\end{enumerate}
In fact the concentration of species $A$ is completely defined by
	\begin{equation}
	R_A = n_A (T) \overline{v_A}(T)
	\label{reaction-rate}
	\end{equation}
where $R_A(T)$ is the emission flux of $A$, $n_A (T)$ is the
concentration of species $A$ and $\overline{v_A}(T)$ is the average
velocity of $A$ which is of order of $\sqrt{T/A}$.\ \ In other words,
the outward flux, at equilibrium, is the same as the inward flux.

Thus a direct connection is made between the statistical decay rate
and Fisher's equilibrium description of cluster formation.\ \ Two
consequences follow:
	\begin{enumerate}
		\item At equilibrium, the evaporated particle is
                      replaced by the back flux from the vapor.\ \
                      However, since the back flux is absent in the
                      case of nuclear multifragmentation, this
                      analysis is limited to particles with low
                      emission probability (first chance) and must
                      avoid particles which are emitted with high
                      multiplicity.\ \ This is approximately achieved
                      by eliminating fragments with $Z<4$ from the
                      ensuing analysis.
		\item In the same spirit as above, the pertinent
                      temperature is that of the blob as it evaporates
                      low probability particles.\ \ Thus, rather than
                      worrying about the role of high multiplicity
                      particles and their associated cooling on the
                      energy-temperature relationship, the Fermi gas
                      relation ship $E= a T^2$ can be assumed with
                      good confidence.
	\end{enumerate}
With this picture in mind, we return to Fisher's formalism.

The basic idea is that in a non-ideal vapor of particles interacting
with repulsive cores and short range attractive forces, can be
approximated at low densities and temperature by an ideal gas
consisting of noninteracting monomers, dimers, trimers at
equilibrium.\ \ The (free) energy of sufficiently large clusters can
be estimated in terms of their volume and surface energy.\ \ These
clusters are in equilibrium with each other and the relative
abundances of differently sized clusters changes with temperature and
pressure \cite{fisher-69}.

The relative abundances of clusters with $A$ constituents is given by:
	\begin{equation}
	n_A = q_0 A^{-\tau} 
	\exp \left[ - \frac{\Delta \mu A}{T} - \frac{c_0 \varepsilon A^{\sigma}}{T} \right] .
	\label{droplet_distribution}
	\end{equation}
Here $q_0$ is a constant of proportionality which is fixed by the
critical density \cite{binder-74,nakanishi-80}.\ \ The power law
$A^{-\tau}$ arises from a combinatorial factor that depends on the
fact that the surface of the cluster must be closed
\cite{fisher-59,essam-63}.\ \ The distance from coexistence is:
	\begin{equation}
	\Delta \mu = {\mu}_l - \mu ,
	\label{delta-mu}
	\end{equation}
where ${\mu}_l$ is the chemical potential the liquid at coexistence
and $\mu$ is the chemical potential of the system.\ \ For $\Delta \mu
> 0$ (a super-heated vapor) and $\Delta \mu = 0$ (liquid-vapor
coexistence) the above sum always converges.\ \ While for $\Delta \mu
< 0$ (a super-saturated vapor), the sum diverges.\ \ The ``classical''
part of the surface energy is parameterized by $c_0 \varepsilon
A^{\sigma}$, where $c_0$ is the zero temperature surface energy
coefficient, $\varepsilon = (T_c - T) / T_c$ and $A^{\sigma}$ relates
the number of constituents of a cluster to the most probable surface
area.\ \ Fisher's critical exponents $\sigma$ and $\tau$ depend on the
Euclidean dimensionality and universality class of the system.

The total pressure of the entire cluster distribution is given by
summing all of the partial pressures $T n_A$:
        \begin{equation}
        p = {\sum} T n_{A}
        \label{pressure}
        \end{equation}
and the density is
        \begin{equation}
        \rho = {\sum} A n_A .
        \label{density}
        \end{equation}
Thus the pressure and density of the system can be inferred from the
knowledge of the cluster distributions.

At the critical point the system is at coexistence ($\Delta \mu = 0$)
and the classical part of the surface energy cost vanishes
(${\varepsilon} = 0$).\ \ Thus both exponential factors are unity
leaving only the temperature independent power law:
        \begin{equation}
        n_{A} = q_{0} A^{-\tau}.
        \label{power_law}
        \end{equation}
Away from the critical point, but along the coexistence curve so that
$\Delta \mu = 0$, the cluster distribution is given by:
        \begin{equation}
        n_{A} = q_{0} A^{-\tau} \exp \left( - \frac{c_0
        \varepsilon A^{\sigma} }{ T } \right) . 
        \label{droplet_distribution_at_mucoex}
        \end{equation}
Equation~(\ref{droplet_distribution_at_mucoex}) can be rewritten as:
        \begin{eqnarray}
        n_{A} & = & \left( q_0 A^{-\tau} \exp \left(
        \frac{c_0 A^{\sigma}}{T_c} \right) \right) \exp \left( -
        \frac{c_0 A^{\sigma}}{T} \right) \nonumber \\ 
	& = & R \exp \left( - \frac{B}{T} \right). 
        \label{boltzmann_distribution}
        \end{eqnarray}
Thus the cluster distribution along the coexistence curve is given by
a Boltzmann factor with
        \begin{equation}
        R = q_0 A^{-\tau} \exp \left( \frac{c_0 A^{\sigma}}{T_c}
        \right)
        \label{prefactor}
        \end{equation}
and
        \begin{equation}
        B = c_0 A^{\sigma} .
        \label{barrier}
        \end{equation}
This Boltzmann factor manifests itself in Arrhenius plots for the
fragment yields where a linear relation between $\ln (n_A )$ and and
$1/T$ is observed.\ \ This behavior has long been observed in many
nuclear fragment yield distributions
\cite{moretto-97,elliott-00.2,tso-95,phair-96,beaulieu-98,moretto-99,beaulieu-01}
and has recently been observed in the cluster distributions of
percolation (with the bond breaking probability playing the role of
temperature) \cite{elliott-00.2} and the Ising model \cite{mader-01}.

As discussed above, Fisher's formalism relates directly to a reaction rate
picture.\ \ In this picture, the heavy fragments (e.g. $Z \ge 4$) are
the product of first chance emission from the excited remnant.\ \ The
first chance emission from a compound nucleus can be written as:
	\begin{equation}
	n_A (T) \propto \Gamma \propto e^{(-B/T)} .
	\label{emission}
	\end{equation}
Thus the fragment yields, parameterized via Fisher, can be related to
the decay rates (widths $\Gamma$).\ \ Furthermore, $\Gamma$, which
controls the first chance emission yields, is the same decay width
which controls the mean emission times $t$ since:
	\begin{equation}
	\Gamma t \approx \hbar ,
	\label{rate}
	\end{equation}
thus
	\begin{equation}
	t \propto \frac{1}{\Gamma} \propto \frac{1}{n_A (T)} \propto e^{(B/T)} .
	\label{rate-relation}
	\end{equation}
The mean time for fragment emission reported by the ISiS Collaboration
\cite{beaulieu-01,beaulieu-00} is well described as a Boltzmann
factor.\ \ It was also noted that the Boltzmann factor describing the
emission times is the same as that describing the fragment yields
\cite{phair-02.1}.\ \ This indicates that the thermal reaction rate
picture is valid for multifragmentation; fragments can be viewed as
being the result of the evaporation of an excited nucleus.

\subsection{Fitting the data}
\label{fitting-the-data}

Preliminary fits of the gold, lanthanum and krypton data with
Eq.~(\ref{droplet_distribution}) led to puzzlingly large results for
$\Delta \mu$ ($\left< \Delta \mu \right> \approx 3$ AMeV) which could
be interpreted as a substantial degree of super-saturation.\ \ A much
more plausible alternative explanation is the lack of account of the
Coulomb effects in Fisher's formalism.\ \
Equations~(\ref{boltzmann_distribution}) and (\ref{barrier}) 
support the presence of a barrier controlling the flux from liquid to
vapor and vice versa.\ \ This barrier should depend not only on the
surface energy of the fragment but should reflect the entire energy
necessary to remove a fragment from the liquid and place it into the
vapor.\ \ At the least, the energy necessary to relocate a charge $Z$
from the bulk to ``near'' the surface of the ``residual'' nucleus
should be evaluated.\ \ This energy is negative and counteracts the
effects of the surface energy.\ \ Furthermore, it is to leading order,
linear in $Z$ and thus in $A$.\ \ Thus the large values of $\Delta
\mu$.\ \ An attempt to include the Coulomb energy explicitly is then
        \begin{equation}
        n_A  = q_0 A^{-\tau} \exp \left( \frac{A \Delta \mu 
             + E_{Coul}}{T} 
             - \frac{c_0 \varepsilon A^{\sigma} }{ T } \right) .
        \label{fisher-coulomb}
        \end{equation}
with 
        \begin{equation}
        E_{\text{Coul}} = \frac{\left( Z_0 - Z \right) Z }
                          { r_0 \left( \left( A_0 - A \right) 
                          ^{1/3} + A^{1/3} \right) } 
                          \left( 1 - e^{ - x \varepsilon } \right)
        \label{coulomb}
        \end{equation}
with $r_0 = 1.2$ fm.\ \ In Eq.~(\ref{coulomb}) there is a recognizable
Coulomb interaction energy of two touching spheres modified by a
factor $(1-\exp(-x \varepsilon))$.\ \ The parameter $x$ (left as a fit
parameter) takes into account the numerical coefficients of the linear
term in $Z$ plus polarization effects, and $\varepsilon$ takes care of
the need for the vanishing difference between the liquid and vapor
near the critical point.\ \ Note that the Coulomb energy discussed in
Eq.~(\ref{coulomb-component}) is different from the Coulomb energy
discussed in Eq.~(\ref{coulomb}).\ \
Equation~(\ref{coulomb-component}) describes the total Coulomb energy
present in the fragmentation process, while Eq.~(\ref{coulomb})
describes the cost in moving a fragment from the nuclear liquid to the
nuclear vapor.

The mass of a fragment $A$ prior to secondary decay was estimated by
multiplying the measured fragment charge $Z$ by $2$ and then by a
factor of $(1+y(E^*/B_f))$ where $B_f$ is the binding energy of the
fragment and $y$ is a fit parameter to allow for an increase or
decrease in the amount of secondary decay.

The temperature was determined by assuming a degenerate Fermi gas,
	\begin{equation}
	 T = \sqrt{\alpha E^*} .
	\label{fermi-gas-temp}
	\end{equation}
The parameter $\alpha$ was taken to be \cite{raduta-97}:
	\begin{equation}
	\alpha = 8 \left( 1 + \frac{E^*}{B_0} \right)
	\label{level-density}
	\end{equation}
in order to accommodate the empirically observed change in $\alpha$
with excitation energy \cite{hagel-88}.\ \ Here $B_0$ is the binding
energy of the fragmenting system.\ \ Using the Fermi gas approximation
to relate $E^*$ and $T$ gives a reasonable estimate of the temperature
of the excited remnant at the time of first chance emission.\ \ It has
been observed that even the isotope ratio thermometer follows the
Fermi gas approximation quite well so long as the average number of
IMFs is less than one \cite{natowitz-01}.

To obtain the concentration of fragments of a given mass, the total
number of fragments $N_A$ of a given size $A$ was normalized to the
size of the fragmenting system $A_0$ so that $n_A = N_A / A_0$.

The location of the critical point, in terms of excitation energy,
was determined from an examination of measured fluctuations.\ \ In
general, as the critical point of a system is approached from the two
phase region, the difference  between phases diminishes and the system
fluctuates from one phase to the other.\ \ At the critical point the
fluctuations are maximal.\ \ However, while the maximum in the
fluctuations occurs at the critical point, the presence of a peak
in the fluctuations is a necessary, but not sufficient, condition for
a existence of a phase transition \cite{elliott-00.1}.

\begin{figure} [ht]
        \centerline{\psfig{file=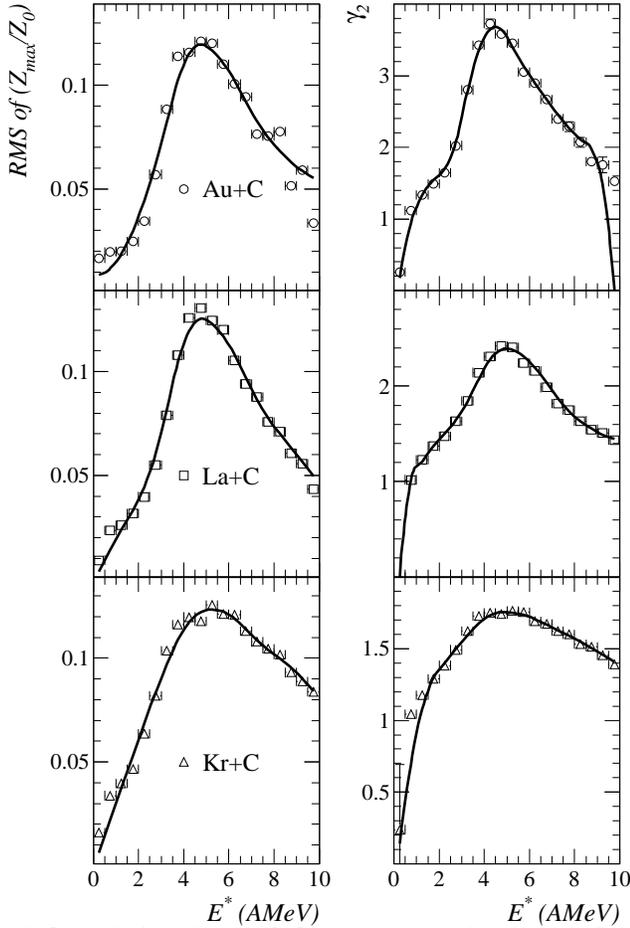,width=8.3cm,angle=0}}
        \caption{Left: The RMS fluctuations in the charge of the
        largest fragment normalized to the charge of the fragmenting
        system plotted as a function of excitation energy.\ \ Right:
        The quantity ${\gamma}_2$ plotted as a function of the
        excitation energy.\ \ Open symbols show the data points, solid
        curves show the results of smoothing the data.}
        \label{fluctuations1}
        \end{figure}

The fluctuations measured in the EOS data are: (1) in the charge of
the largest fragment normalized to the charge of the fragmenting
system, and (2) related to the average mass number of a fragment as
measured by the quantity ${\gamma}_2$ \cite{campi-88}, where
	\begin{equation}
	{\gamma}_2 = \left( \frac{\text{RMS}(A)}{\left< A \right> } \right) ^2 + 1 = \frac{M_2 M_0}{M_{1}^{2}}
	\label{gamma-2}
	\end{equation}
with $M_k$ as the $k^{\text{th}}$ moments of the fragment distributions
	\begin{equation}
	M_k = \sum_{A=1} n_A A^k .
	\label{moments}
	\end{equation}
These fluctuations are shown in Fig.~\ref{fluctuations1}.

\begin{figure} [ht]
        \centerline{\psfig{file=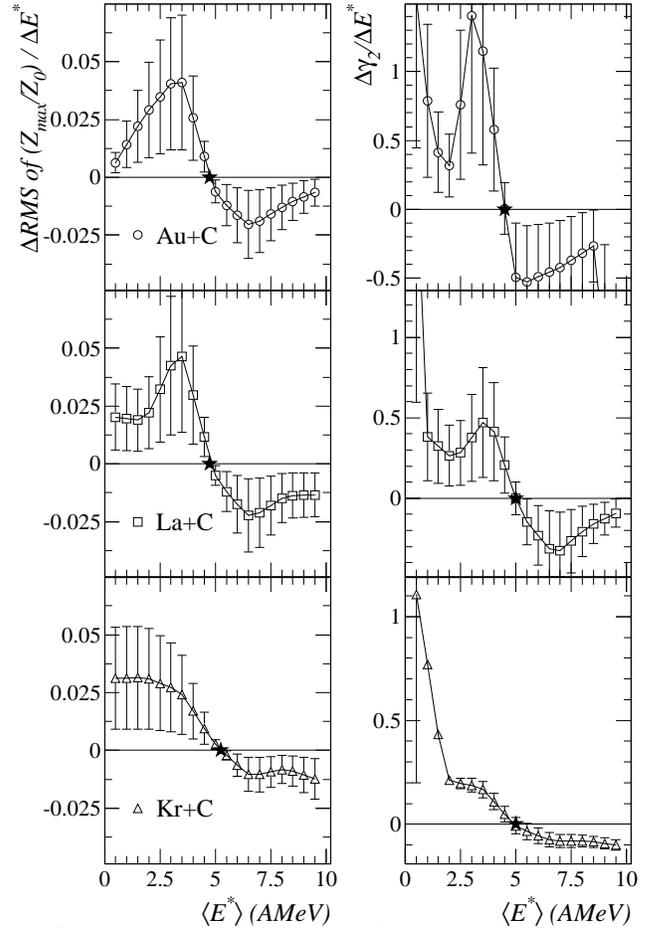,width=8.3cm,angle=0}}
        \caption{Numerical derivatives of the smoothed data from
        Fig.~\ref{fluctuations1} of: (left) the RMS fluctuations in the
        charge of the largest fragment normalized to the charge of the
        fragmenting system plotted as a function of excitation
        energy, and (right) the quantity ${\gamma}_2$ plotted as a
        function of the excitation energy.\ \ Solid star symbols
        shows the approximate location in excitation energy where the
        derivative is zero, thus indicating the critical point.}
        \label{fluctuations2}
        \end{figure}

The peak in the fluctuations was found by smoothing the data (solid
lines in Fig.~\ref{fluctuations1}), taking the numerical derivative of
the smoothed data and finding the value of $E^*$ where the the
derivative passed through zero, see Fig.~\ref{fluctuations2}.\ \
Finally, the value of the excitation energy at the critical point
$E^{*}_{c}$ was determined by averaging the results from both measures
of the fluctuations.\ \ Table~\ref{critical-point-table} lists the
results.\ \ For this analysis the values determined for the excitation
energy at the critical point for the Au$+$C reaction are in proximity
of other values observed in previous EOS analyses ($E^{*}_{c} \approx
4.75$ AMeV)
\cite{elliott-00.2,hauger-98,hauger-00}.\ \ Differences
in the values of $E^{*}_{c}$ arise from the different method of
constructing the thermal portion of the excitation energy described
above \cite{srivastava-02,srivastava-01}.

Estimates of the critical temperature $T_c$ are made by using the
values of $E^{*}_{c}$ in Eq.~(\ref{fermi-gas-temp}) and lead to
values, shown in Table~\ref{critical-point-table}, that are comparable
to theoretical estimates for small nuclear systems
\cite{jaqaman-83,jaqaman-84,bonche-85,de-99}.\ \ As an aside, as shown
in Table~\ref{critical-point-table} the value of $T_c$ increases with
decreasing projectile (and thus remnant) mass.\ \ This is opposite of
the trend assumed in a prior analysis of the EOS gold
multifragmentation data where the Coulomb energy was neglected
\cite{elliott-00.1} but in agreement with the trend reported in other
analysis of the EOS data sets \cite{srivastava-02,srivastava-01}.

In Fig.~\ref{fluctuations1} the value of ${\gamma}_2$ for the Kr
system attains a peak value of only ${\sim}1.8$.\ \ It has been
suggested that the magnitude of the peak in ${\gamma}_2$ could
distinguish between the presence of a power law with $\tau > 2$
(${\gamma}_2 > 2$) and an exponential distribution (${\gamma}_2 < 2$)
in the cluster yields \cite{srivastava-01,campi-88}.\ \ However, it
was seen that this is not the case \cite{elliott-00.1} and it
will be seen in the Appendix that small percolation lattices have
values of ${\gamma}_2$ with peak magnitudes of less than two yet still
exhibit a continuous phase transition with an exponent of $\tau \sim
2.2$ in the power law describing the cluster yields at the critical
point.\ \ Thus, the height of the peak in ${\gamma}_2$ cannot be used
to rule out the presence of a critical point and the associated power
law in the cluster distribution or provide information about the value
of the power law exponent.

	\begin{table}
	\caption{Critical points of excited nuclei}
	\begin{tabular}{lllll}
	System   & $E^{*}_{c}$ (AMeV) & $T_c$ (MeV)   & ${\rho}_c$ (${\rho}_0$) & $p_c$ (MeV$/$fm$^3$) \\
	\hline
	Au $+$ C & $4.6 \pm 0.2$      & $7.6 \pm 0.2$ & $0.39 \pm 0.01$        & $0.11 \pm 0.04$       \\
	La $+$ C & $4.9 \pm 0.2$      & $7.8 \pm 0.2$ & $0.39 \pm 0.01$        & $0.12 \pm 0.04$       \\
	Kr $+$ C & $5.1 \pm 0.2$      & $8.1 \pm 0.2$ & $0.39 \pm 0.01$        & $0.12 \pm 0.04$       \\
	\end{tabular}
	\label{critical-point-table}
	\end{table}

Data from each system for $0.25$ AMeV $\le E^* \le E^{*}_{c}$ (which
corresponds to a range of $0 \le \varepsilon < \sim 0.8$) and $4 \le Z
\le Z_0/4$ were simultaneously fit to Eq.~(\ref{fisher-coulomb}),
which, as mentioned previously, helps insure that the fragments
examined in this analysis are produced via first chance emission.\ \
There were nearly $200$ points from the EOS data sets used in the
fitting procedure.\ \ The fit parameters $\tau$, $\sigma$ and $c_0$
were kept the same for all three data sets while ${\Delta \mu}$, $x$
and $y$ were allowed to vary between the systems to minimize
chi-squared; this gives $12$ free parameters used to fit nearly $200$
data points.\ \ Previous analyses of the EOS data
\cite{elliott-00.1,elliott-00.2} assumed that $\Delta \mu = 0$ and
that the effects of the Coulomb energy were small.\ \ The analysis
presented here makes no such assumptions.

Fixing $\tau$ at 2.2 did not significantly change the results of this
analysis.\ \ Using a common $x$ value for all three data sets also
returned results similar to those quoted below.\ \ Using a common $y$
value for all three data sets also returned results similar to those
quoted below. \ \ These different methods suggest a systematic error
of $\sim15$\% of the value in question.\ \ All errors quoted below are
those returned by the fitting procedure, propagated where necessary.\
\ Finally, the same data collapse observed below would be seen if the
parameters were fixed to: $\tau = 2.21$, $\sigma = 0.64$ (their $d=3$
Ising values), $c_0 = 16.8$ MeV (the text book value of the nuclear
liquid-drop surface energy coefficient), ${\Delta
\mu}_{\text{Au,La,Kr}}=0$ (they must be close to zero since fragments
are observed) and $y=0.5$ (in keeping with previous assumptions that
the fragments prior to secondary decay have the same mass to charge
ratio of the excited remnant \cite{elliott-00.1,srivastava-02}) and
letting only $x$, the Coulomb parameter vary to minimize chi-squared.

\subsection{Results}
\label{Results}

        \begin{figure*}
        \centerline{\psfig{file=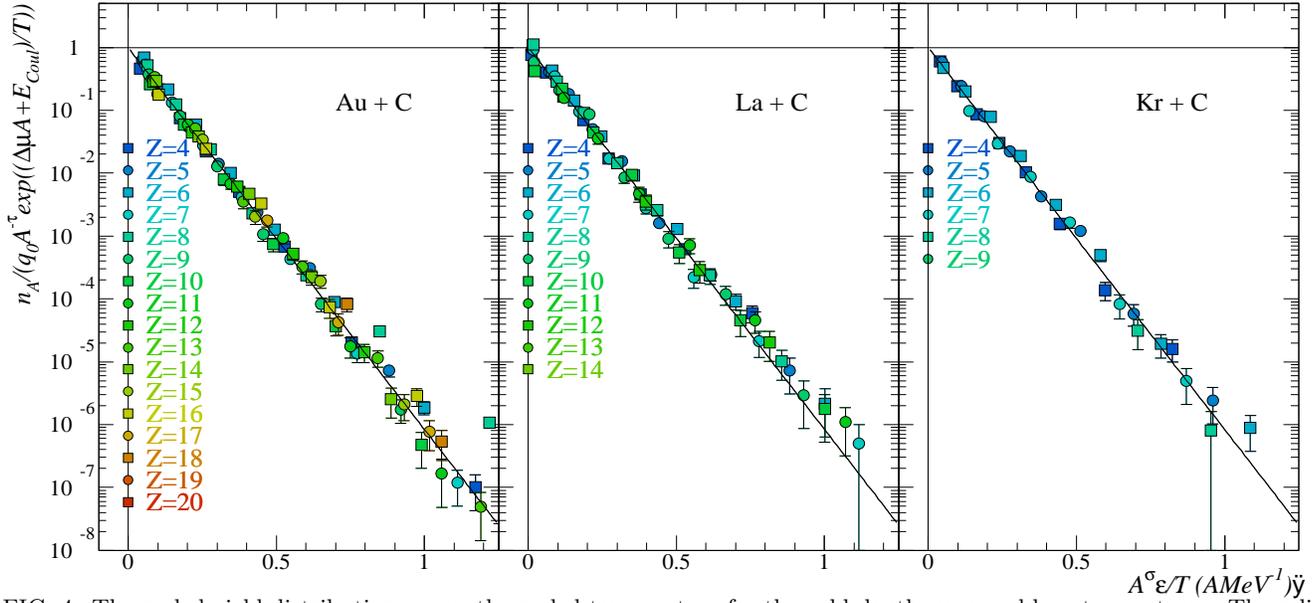,width=17.2cm,angle=0}}
        \caption{The scaled yield distribution versus the scaled
        temperature for the gold, lanthanum and krypton systems.\ \
        The solid line has a slope of $c_0$.}
        \label{scaling}
        \end{figure*}

Figure~\ref{scaling} shows the fragment mass yield distribution scaled
by the power law pre-factor, the chemical potential and
Coulomb terms: $n_A / q_0 A^{-\tau} \exp(({\Delta}{\mu} A +
E_{\text{Coul}}) / T)$ plotted against the inverse temperature scaled
by Fisher's parameterization of the surface energy: $A^{\sigma}
\varepsilon / T$.\ \ Now, the scaled data for all three systems
collapse onto a single line over several orders of magnitude as
predicted by Fisher's droplet formalism \cite{fisher-67}.\ \ This
collapse provides direct evidence for a liquid to vapor phase
transition in excited nuclei.\ \ Furthermore, the fact that the data
from each system show a common scaling illustrates the common nature
of the underlying phenomenon.

\subsubsection{Parameters}
\label{Parameters}

        \begin{table}
        \caption{Uncommon fit parameters}
        \begin{tabular}{llll} 
         System  & $\Delta \mu$ (AMeV) & $x$             & $y$           \\
        \hline
        Au $+$ C & $0.38 \pm 0.02$     & $1.1 \pm 0.2$   & $0.5 \pm 0.1$ \\
        La $+$ C & $0.47 \pm 0.03$     & $1.2 \pm 0.1$   & $0.3 \pm 0.2$ \\
        Kr $+$ C & $0.58 \pm 0.08$     & $4.0 \pm 1.0$   & $0.8 \pm 0.2$ \\
        \end{tabular}
        \label{uc-param-table}
        \end{table}

The values of $\tau = 2.2 \pm 0.1$, $\sigma = 0.71 \pm 0.02$ and $c_0
= 14.0 \pm 1.0$ MeV determined in this analysis are in agreement with
those determined for the ISiS gold multifragmentation data sets
\cite{elliott-02} and are in agreement with values previously
determined for the EOS Au$+$C data set
\cite{srivastava-02,elliott-00.2}.\ \ The value of the surface energy
coefficient $c_0$ is close to the value of the surface energy
coefficient of the liquid-drop model which is $\sim 16.8$ MeV.

A previous analysis of the EOS gold multifragmentation showed the
surface energy coefficient to be $c_0 = 6.8 \pm 0.5$ MeV
\cite{elliott-00.2}.\ \ The difference between the $c_0 = 6.8 \pm 0.5$
MeV from that work and the $c_0 = 14.0 \pm 1.0$ MeV presented here
arises from the differing analyses.\ \ In the previous analysis it was
assumed that $\Delta \mu = 0$, that the Coulomb energy was negligible
and that the level density parameter was constant at $\alpha = 13$.\ \
These assumptions allowed some degree of scaling, yielded sensible
values for the critical exponents but resulted in a surface energy
coefficient that was a factor of two of lower than that of the present
analysis.

In addition to a surface energy coefficient that is in better
agreement with the standard liquid-drop model, the greater collapse of
the data in the present work demonstrates the improvements of the
present analysis over the previous one.\ \ The improvements in analysis
are related to allowing a non-zero $\Delta \mu$, taking into account the cost in
Coulomb energy to move a fragment from the liquid to the vapor and
accounting for the change in the level density parameter over the
excitation energy range.\ \ The treatment of secondary decay in both
analyses is different: previously it was assumed that the fragments,
prior to any secondary decay, had the same mass to charge ratio as the
fragmenting remnant.\ \ In the present analysis the amount of
secondary decay is left as a free parameter.

The values of $\Delta \mu$ reported in Table~\ref{uc-param-table} can
be considered ``small'' in light of Eq.~(\ref{delta-mu}).\ \ The
chemical potential of the liquid can be found by 
	\begin{equation}
	{\mu}_l = E_0 + T S_0
	\label{chem-pot1}
	\end{equation}
with $E_0$ as the bulk energy per particle and $S_0$ as the bulk
entropy per particle \cite{fisher-67}.\ \ Treating the system as a
Fermi gas so that $S/A=\alpha T$ yields
	\begin{equation}
	{\mu} = E_0 + E^* .
	\label{chem-pot2}
	\end{equation}
Thus the important energy scale for $\Delta \mu$ is $E_0 + E^*$, for
nuclear matter $E_0 \sim 15.5 MeV$.\ \ The values of $\Delta \mu$
returned by this analysis are $< 6$\% of $E_0 + E^*$ indicating
the system is close to coexistence.\ \ The values of $\Delta \mu$
should also be compared to the values returned when the EOS fragment
yields were fit to (Eq.~\ref{droplet_distribution}): $\left< \Delta
\mu \right> \approx 3.0$ AMeV for all EOS reactions.\ \ The reduction
in the magnitude of the $\Delta \mu$ values is about a factor of six
and is due to the modification of Eq.~(\ref{droplet_distribution}) to
account for the Coulomb energy, i.e. Eq.~(\ref{fisher-coulomb}).\ \
The remaining small positive $\Delta \mu$ values of the systems may
indicate that those systems are slightly super-saturated, or more
probably they may reflect some other energy costs not taken into
account (e.g. the symmetry energy or pairing), or they may reflect
that the approximation for the cost in Coulomb energy to form a
fragment given in Eq.~(\ref{coulomb}) is not completely adequate (for
instance Eq.~(\ref{coulomb}) assumes a spherical geometry which may or
may not be the case), or they may merely reflect noise in the data.

The values of $x$ for each system may indicate more (Au and La) or
less (Kr) Coulomb energy present in the system.\ \ They may also
reflect the symmetry of the collision which may affect the geometry of
the remnant, e.g. a very asymmetric collisions like Au$+$C may leave a
nearly spherical remnant, while a more symmetric collision like Kr$+$C
may result in a less spherical fragmenting system.

The values of $y$ returned indicate that the fragments have the same
mass to charge ratio as the excited remnant.

The difference in values of $\Delta \mu$, $x$ and $y$ determined in
the analysis of the three EOS data sets and those determined in the
analysis of the ISiS $8.0$ GeV$/$c $\pi$ on gold multifragmentation
set \cite{elliott-02} is left an open question.\ \ The small
differences in $E^{*}_{c}$ and $T_c$ are due to the differences in
reconstructed excitation energy scales \cite{lefort-99}.\ \ This
difference carries over to all energy related quantities, e.g. $c_0$.

Finally, in light of the above parameter results, it is clear that the
same data collapse would be observed if the parameters were fixed to
some nominal values, discussed above, with only $x$, the Coulomb
parameter varying to minimize chi-squared.\ \ Thus only three free
parameters are truly needed to fit the $\sim 200$ data points of the
EOS data sets.

\subsection{The coexistence curve of finite neutral nuclear matter}
\label{phase-diagram}

\subsubsection{The pressure-temperature coexistence line}
\label{pt-coexistence-line}

Before determining the pressure-temperature coexistence line, the
meaning of a pressure associated with an excited nuclear remnant must
be addressed.\ \ As discussed above, in the actual experiment, this
pressure is virtual; it is the pressure the vapor would have in order
to provide the back flow needed to keep the source at equilibrium.\ \
However, since the yields from Fisher's formalism are proportional to both
the pressure, Eq.~(\ref{pressure}), and the evaporation rate,
Eq.~(\ref{rate-relation}), it is clear that by fitting the yields as
has been done above, one can infer an associated (virtual) vapor
pressure.

        \begin{figure} [ht]
        \centerline{\psfig{file=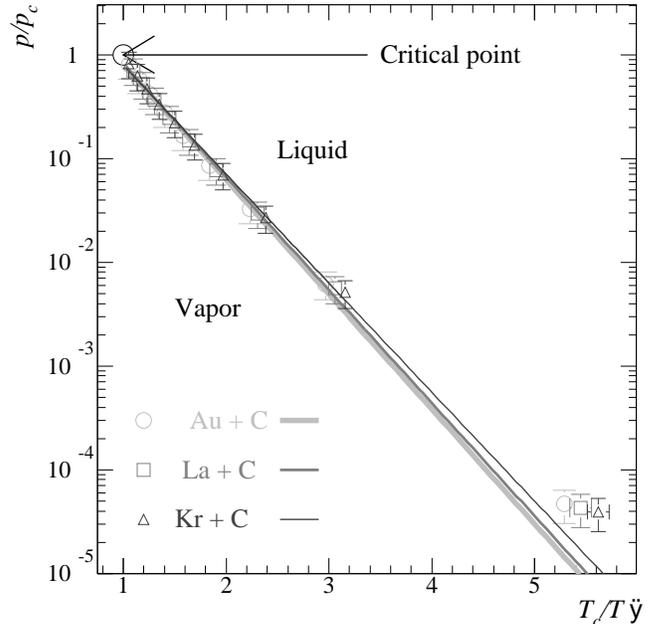,width=8.3cm,angle=0}}
        \caption{The reduced pressure-temperature phase diagram: the
        points show calculations performed at the excitation energies
        below the critical point and the lines show fits to the
        Clausius-Clapeyron equation.}
        \label{temp-pres}
        \end{figure}

The $p$-$T$ coexistence curve can be determined from this analysis.\ \
As seen in section~\ref{Fisher's-droplet-formalism}, Fisher's theory
assumes that the non-ideal fluid can be approximated by an ideal gas
of clusters.\ \ Accordingly, the quantity $n_A$ is proportional to the
partial pressure of a fragment of mass $A$ and the total pressure due
to all of the fragments is the sum of their partial pressures (see
Eq.~(\ref{pressure})).\ \ The reduced pressure is then given by:
        \begin{equation}
        \frac{p}{p_c} = \frac{T   \sum n_A(T)  }
                             {T_c \sum n_A(T_c)}.  
        \label{reduced_pressure}
        \end{equation}
The coexistence curve for finite neutral nuclear matter is obtained by
substituting the $n_A(T,\Delta \mu = 0, E_{Coul} =0)$ from
Eq.~(\ref{fisher-coulomb}) in the numerator of
Eq.~(\ref{reduced_pressure}) and $n_A(T_c,\Delta \mu = 0, E_{Coul}
=0)$ in the denominator.\ \ This allows one to transform the
information in Fig.~\ref{scaling} into the familiar phase diagram in
Fig.~\ref{temp-pres}.\ \ The data points shown give the values of
$p/p_c$ and $T_c/T$ calculated via Eq.~(\ref{reduced_pressure}) for
the bins in $E^*$ up to and including the critical point.

Figure~\ref{temp-pres} gives an estimate of the coexistence line of
finite nuclear matter and from this it is possible to make an estimate
of the bulk binding energy of nuclear matter.\ \ One begins by
assuming the system behaves as an ideal gas and uses the
Clausius-Clapeyron equation 
	\begin{equation}
	\frac{\partial p}{\partial T} = \frac{\Delta H}{T \Delta V}
	\label{claus-clap}
	\end{equation}
where $\Delta H$ is the molar enthalpy of evaporation and $\Delta V$
is the molar volume difference between the two phases. \ \ Then solving
for the vapor pressure with
	\begin{equation}
	\Delta V = V_{\text{vapor}} - V_{\text{liquid}} \approx V_{\text{vapor}} = \frac{T}{p}
	\end{equation}
gives
	\begin{equation}
	p = p_0 \exp(\frac{-\Delta H}{T})
	\label{pressure-exp}
	\end{equation}
which would lead to the ratio of
	\begin{equation}
	\frac{p}{p_c} = \exp \left[ \frac{\Delta H}{T_c} \left(1 - \frac{T_c}{T} \right) \right] .
	\label{gugg-01}
	\end{equation}
if $\Delta H$ were assumed to be temperature independent.\ \ However,
as $T \rightarrow T_c$ the gas is not ideal and $\Delta H \neq
\text{constant}$, but it has long been known that for several normal
fluids these deviations compensate so that $\ln (p/p_c)$ is
approximately linear in $T/T_c$ \cite{guggenheim-text}.

A fit of Eq.~(\ref{gugg-01}) to the coexistence curves for the systems
is shown in Fig.\ref{temp-pres} yields the ratio of $\Delta H / T_c$.\
\ Using the corresponding values of $T_c$ gives the molar enthalpies
of evaporation of the liquid $\Delta H$ shown in
Table~\ref{thermo-table}.\ \ From these $\Delta H$ values $\Delta E$
is constructed via $\Delta E = \Delta H - pV$ with $pV=T$ (with the
ideal gas approximation) using the average temperature from the range
in Fig.~\ref{temp-pres} listed in Table~\ref{thermo-table}.\ \ $\Delta
E$ refers to the cost in energy to evaporate a single fragment.\ \ To
determine the energy cost on a per nucleon basis $\Delta E$ is divided
by the most probable size of a fragment over the temperature range in
Fig.~\ref{temp-pres}.\ \ Since the gas described by Fisher's formalism
is an ideal gas of clusters, the most probable cluster size is greater
in size than a monomer.\ \ The most probable size of a fragment in the
region of the $p$-$T$ coexistence line obtained from
Eq.~(\ref{fisher-coulomb}) and the experimentally determined
parameters is $1.05 \pm 0.05$.\ \ Thus the $\Delta E / A$ becomes
$\approx$14 AMeV, close to the nuclear bulk energy coefficient of
$15.5$ MeV.

	\begin{table}
	\caption{Thermodynamic properties of excited nuclei}
	\begin{tabular}{lllll}
	System   & $\Delta H$ (MeV) & $\left< T \right> $ (MeV) & $\Delta E / A$ (AMeV) & $C^{F}_{c}$    \\
	\hline
	Au $+$ C & $19.4 \pm 0.7$   & $4.6 \pm 0.6$             & $14 \pm 1$            & $0.28 \pm 0.09$ \\
	La $+$ C & $19.6 \pm 0.7$   & $4.9 \pm 0.6$             & $14 \pm 1$            & $0.28 \pm 0.09$ \\
	Kr $+$ C & $19.5 \pm 0.7$   & $4.9 \pm 0.6$             & $14 \pm 1$            & $0.28 \pm 0.09$ \\
	\end{tabular}
	\label{thermo-table}
	\end{table}

\subsubsection{The temperature-density coexistence curve}
\label{rt-coexistence-curve}

        \begin{figure} [ht]
        \centerline{\psfig{file=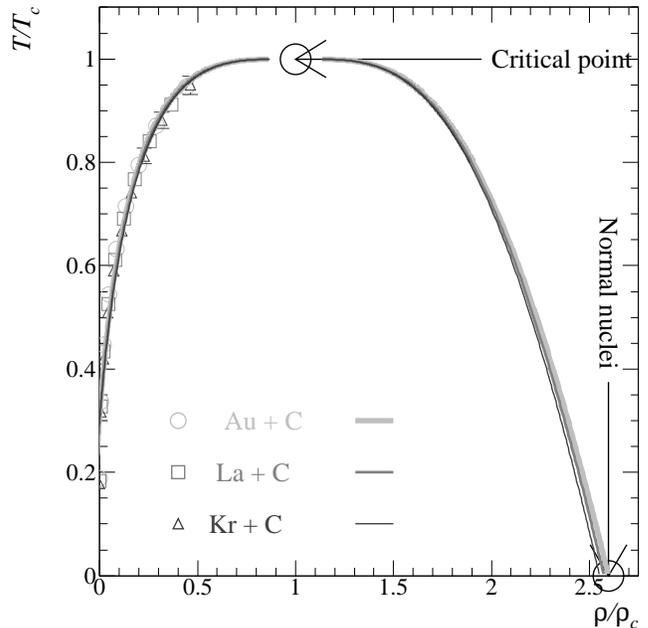,width=8.3cm,angle=0}}
        \caption{The points are calculations performed at the
        excitation energies below the critical point and the lines are
        a fit to and reflection of Guggenheim's equation.}
        \label{temp-den}
        \end{figure}

As seen in section~\ref{Fisher's-droplet-formalism} the system's
density can be found from Eq.~(\ref{density}).\ \ The reduced density is
given by:
        \begin{equation}
        \frac{\rho}{{\rho}_c} = \frac{\sum A n_A(T)  }
                                     {\sum A n_A(T_c)} .
        \label{reduced_density}
        \end{equation}
With $\Delta \mu$ and $E_{Coul}$ set to $0$ in the numerator of
Eq.~(\ref{fisher-coulomb}) and $\Delta \mu$ and $E_{Coul}$ set to $0$
with $T$ set to $T_c$ in the denominator, Eq.~(\ref{reduced_density})
gives the low density (vapor) branch of the coexistence curve of finite
nuclear matter, shown in Fig.~\ref{temp-den}.

Following Guggenheim's work with simple fluids, it is possible to
determine the high density (liquid) branch as well: empirically, the
$\rho/{\rho}_c$-$T/T_c$ coexistence curves of several fluids can be
fit with the function \cite{guggenheim-45}:
	\begin{equation}
        \frac{{\rho}_{l,v}}{{\rho}_c} = 1 + b_1 (1-\frac{T}{T_c}) 
        \pm b_{2} (1-\frac{T}{T_c})^{\beta}
        \label{gugg-eq}
        \end{equation}
where the parameter $b_2$ is positive (negative) for the liquid
${\rho}_l$ (vapor ${\rho}_v$) branch.\ \ Using Fisher's formalism, $\beta$
can be determined from $\tau$ and $\sigma$ \cite{fisher-67}:
	\begin{equation}
	\beta = \frac{\tau - 2}{\sigma} .
	\label{beta}
	\end{equation}
For this work $\beta = 0.3 \pm 0.1$.\ \ Using this value of $\beta$
and fitting the coexistence curve from the EOS data sets with
Eq.~(\ref{gugg-eq}) one obtains estimates of the ${\rho}_v$ branch of
the coexistence curve and changing the sign of $b_{2}$ gives the
${\rho}_l$ branch, thus yielding the full $T$-$\rho$ coexistence curve
of finite nuclear matter.

From Fig.~\ref{temp-den} it is possible to make an estimate of the
density at the critical point ${\rho}_c$.\ \ Assuming that normal
nuclei exist at the $T=0$ point of the ${\rho}_l$ branch of the
coexistence curve, then using the parameterization of the coexistence
curve in Eq.~(\ref{gugg-eq}) gives ${\rho}_c \sim {\rho}_0/3$.\ \ See
Table~\ref{critical-point-table} for precise values.

\subsubsection{The pressure-density coexistence curve}
\label{pr-coexistence-curve}

        \begin{figure} [ht]
        \centerline{\psfig{file=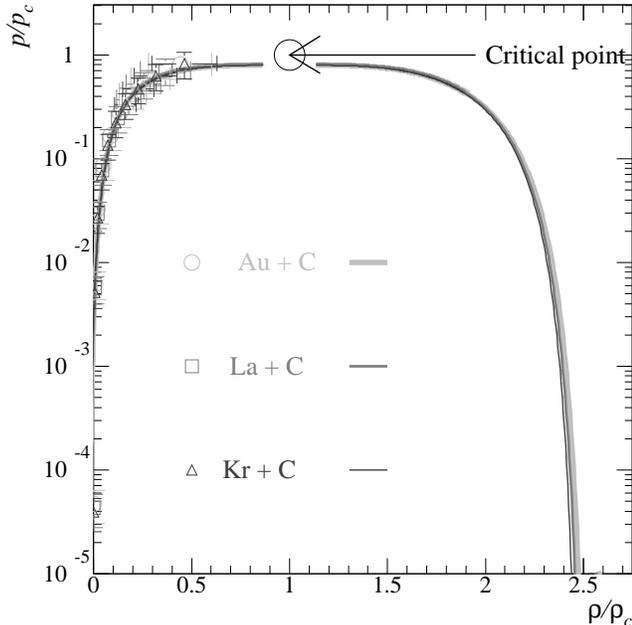,width=8.3cm,angle=0}}
        \caption{The points are calculations performed at the
        excitation energies below the critical point and the lines are
        the results of the fits from the previous sections.}
        \label{pres-den}
        \end{figure}

For the sake of completeness the $p/p_c$-${\rho}/{\rho}_c$ projection
of the coexistence curve is determined by combining the results of the
previous two sections.\ \ This is shown in Fig.~\ref{pres-den}.\ \ It
is clear from Fig.~\ref{pres-den} that the fitted curves do not reach
$p/p_c = 1$ at ${\rho}/{\rho}_c = 1$ while the data points do.\ \ This
is a reflection of the validity of the assumptions that went into
deriving Eq.~(\ref{gugg-01}).

\subsubsection{The compressibility factor}
\label{comp-fact}

The critical compressibility factor $C^{F}_{c} = p_c / T_c {\rho}_c$
can also be determined in a straightforward manner from
\cite{kiang-70}:
	\begin{equation}
	C^{F}_{c} = \frac{\sum n_A ( T_c )}{\sum A n_A ( T_c) }
	\label{comp-fac-eq}
	\end{equation}
Table~\ref{thermo-table} shows the results for the EOS data sets
which are in agreement with the values for several fluids
\cite{kiang-70} and that of the ISiS data \cite{elliott-02}.

Finally, a measure of the pressure at the critical point $p_c$ can be
made by using $T_c$ and ${\rho}_c$ from above in combination with
$C^{F}_{c}$.\ \ The results are shown in
Table~\ref{critical-point-table}.\ \ This last calculation gives a
complete experimental measure of the location of the critical point of
finite neutral nuclear matter ($p_c, T_c, {\rho}_c$) and is in
agreement with the ISiS results and in rough agreement with
theoretical calculations \cite{jaqaman-83,de-99}.

        \begin{figure} [ht]
        \centerline{\psfig{file=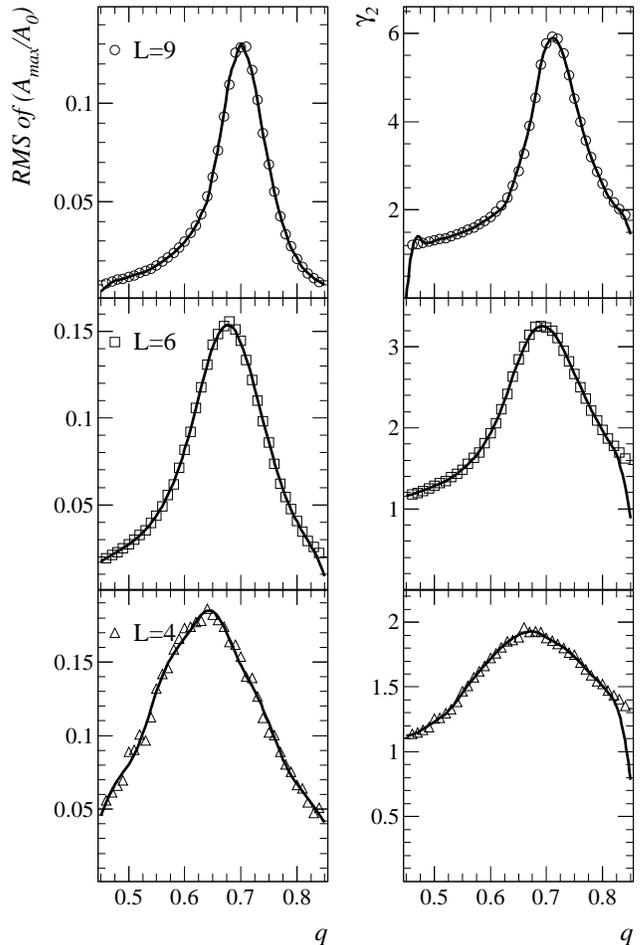,width=8.3cm,angle=0}}
        \caption{Left: The RMS fluctuations in the size of the largest
        cluster normalized to the size of the lattice system plotted as
        a function of bond breaking probability.\ \ Right: The
        quantity ${\gamma}_2$ plotted as a function of bond breaking
        probability.\ \ Open symbols show the estimate of the
        excitation at the critical point based on the maximum of the
        fluctuations, the solid line shows the results of smoothing
        the data.}
        \label{perc-fluc1}
        \end{figure}

        \begin{figure} [ht]
        \centerline{\psfig{file=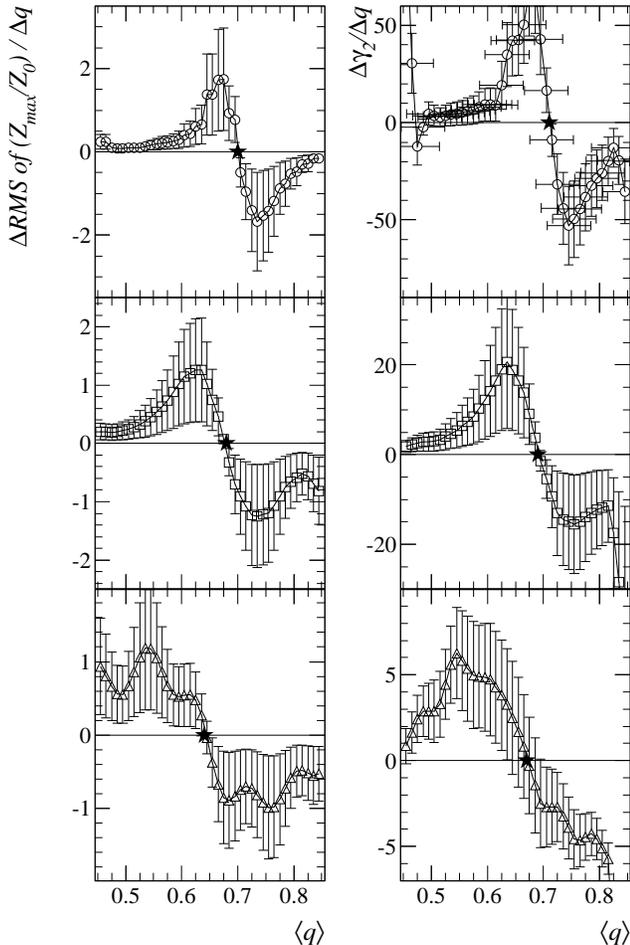,width=8.3cm,angle=0}}
        \caption{Left: The numerical derivative of RMS fluctuations in
        the size of the largest cluster normalized to the size of the
        lattice system plotted as a function of bond breaking
        probability.\ \ Right: The numerical derivative of the
        quantity ${\gamma}_2$ plotted as a function of bond breaking
        probability.\ \ Open symbols show the estimate of the
        excitation at the critical point based on the maximum of the
        fluctuations, the solid stars show where the derivatives are
        zero.}
        \label{perc-fluc2}
        \end{figure}

\section{Conclusion}
\label{Conclusion}

Through a direct examination of the most accessible features of
nuclear multifragmentation, namely the fragment distributions
themselves, and the use of Fisher's droplet formalism, modified to
account for the Coulomb energy cluster formation, a measurement of the
coexistence curve of finite neutral nuclear matter has been made for
three different multifragmenting systems and estimates of the critical
point for finite nuclear matter have been made.\ \ The precise values
of quantities like the critical exponents and critical temperature and
precise locations of coexistence curves depend on the assumptions made
for the cost in Coulomb energy for fragment formation and the
assumptions made to account for the secondary decay of the fragments.\
\ While the exact forms are unknown, the estimates made in this paper
have solid physical origins, and yield values of the surface energy
coefficient and the bulk binding energy of nuclear matter which are
consistent with established values.\ \ Both the $p$-$T$ coexistence
lines and the $T-\rho$ coexistence curves for all three EOS systems
are consistent.\ \ These are strong indications that this analysis
determines the coexistence curve and can be used to construct the
phase diagram of finite neutral nuclear matter {\it based on
experimental data}.

\section{Appendix}

	\begin{table}
	\caption{Critical points of finite percolation lattices}
	\begin{tabular}{llll}
	L   & $q_c$            & ${\rho}_c$         & $p_c$             \\
	\hline
	$9$ & $0.705 \pm 0.004$ & $0.210 \pm 0.001$ & $0.041 \pm 0.001$ \\
	$6$ & $0.685 \pm 0.004$ & $0.216 \pm 0.001$ & $0.041 \pm 0.001$ \\
	$4$ & $0.655 \pm 0.004$ & $0.243 \pm 0.002$ & $0.044 \pm 0.001$ \\
	\end{tabular}
	\label{perc-critical-point}
	\end{table}

To demonstrate the efficacy of the above analysis, it is applied to
the cluster distributions from three dimensional simple cubic lattices
of side $L = 4$, $6$ and $9$.\ \ It will be seen that if the above
procedures are followed, well known quantities are recovered.

Cluster distributions for over $100,000$ lattice realizations were
generated by breaking bonds between sites \cite{elliott-97}.\ \ A
value of the lattice's bond breaking probability $q$ was chosen from a
uniform distribution on (0,1).\ \ Next, a bond probability, $q_i$, was
randomly chosen from a uniform distribution on (0,1) for the
$i^{\text{th}}$ bond.\ \ If $q_i$ was less than $q$, then the
$i^{\text{th}}$ bond was broken and two sites were separated.\ \ This
process was performed for each bond in the lattice.\ \ At low values
of $q$, few bonds were broken resulting in a cluster distributions
that are analogous to the liquid-vapor coexistence of a fluid.\ \ In
an infinite lattice the distinguishablity of the ``liquid'' phase and
the ``vapor'' phase vanishes at a unique value of the lattice
probability, $q_c$, when the probability of forming a percolating
cluster changes from zero to unity \cite{stauffer-79,stauffer-text}.\
\ For the ensuing analysis, the number of clusters of size $A$ per
lattice site $n_A$ was calculated by histogramming the lattice
realizations into 100 bins on $q$ from $0$ to $1$.

        \begin{figure*}
        \centerline{\psfig{file=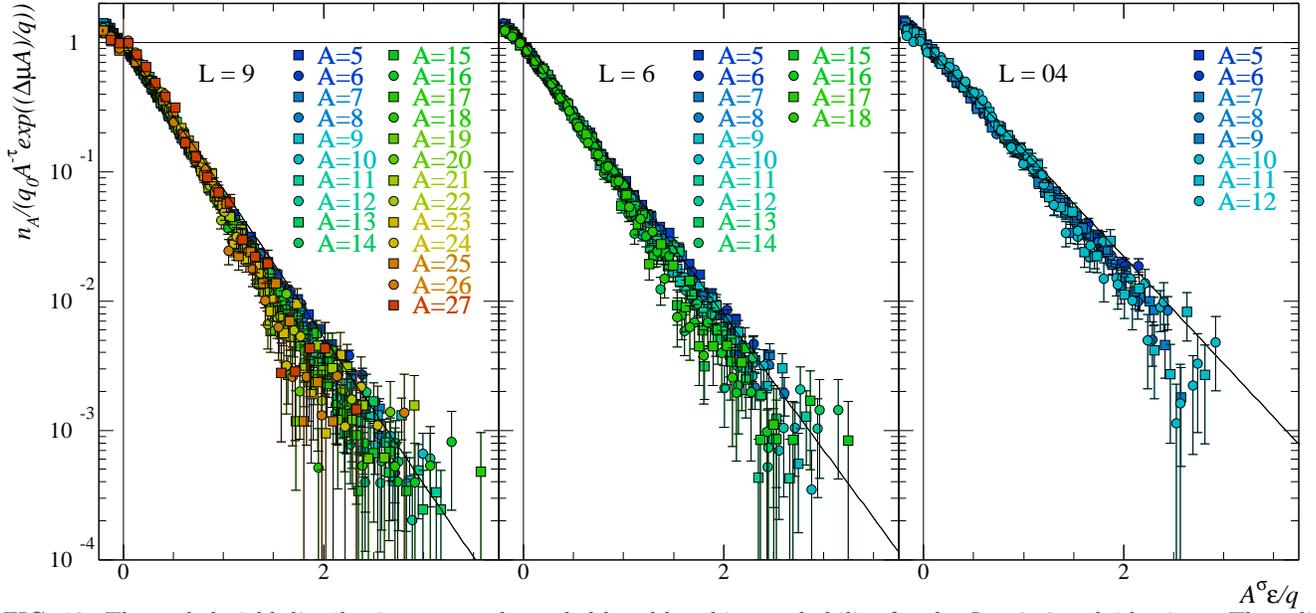,width=17.2cm,angle=0}}
        \caption{The scaled yield distribution versus the scaled bond
        breaking probability for the $L = 9$, $6$ and $4$ lattices.\ \
        The solid lines has a slope of $c_0(L)$.}
        \label{perc-scal}
        \end{figure*}

First the value of the probability at the critical point $q_c$ is
determined by locating the maximum in the fluctuations of: (1) the
size of the largest cluster, and (2) ${\gamma}_2$.\ \
Figures~\ref{perc-fluc1} and \ref{perc-fluc2} show these measures of
the fluctuations.\ \ The location of the maximum is determined as in
the EOS data, the data is smoothed and then the numerical derivative
is taken.\ \ The location of the peak in the largest cluster is
averaged with the location of the peak in ${\gamma}2$ and the results
are recorded in Table~\ref{perc-critical-point}.\ \ As expected the
value of $q_c$ changes with the lattice size.

Note that in Fig.~\ref{perc-fluc1} the value of ${\gamma}_2$ for the
$L=4$ lattice attains a peak value of only ${\sim}1.9$; this is a
finite size effect and due to the small size of the lattice.\ \ Since
${\gamma}_2$ is related to the fluctuations in the average size of a
cluster, it is clear that as the size of the lattice decreases, the
upper limit in the size of a cluster decreases, thus imposing a limit
on the size of ${\gamma}_2$.

        \begin{table}
        \caption{Percolation fit parameters}
        \begin{tabular}{lll} 
        L   & $\Delta \mu$         & $c_0$           \\
        \hline
        $9$ & $-0.008 \pm 0.004$  & $2.62 \pm 0.04$ \\
        $6$ & $0.001  \pm 0.001$  & $2.42 \pm 0.04$ \\
        $4$ & $0.007  \pm 0.001$  & $1.91 \pm 0.04$ \\
        \end{tabular}
        \label{perc-params}
        \end{table}

Next the cluster yields from the three different lattices are fit
simultaneously to Eq.~(\ref{droplet_distribution}), with $q_c(L)$
keeping the fit parameters $\sigma$ and $\tau$ consistent between
lattices and letting $\Delta \mu$ and $c_0$ vary between lattices.\ \
Data from $0.4 \le q \le 1.05q_c$ and $5 \le A \le 3L$ were included
in the fitting procedure.\ \ This gives seven fit parameters with
$1083$ points to fit.\ \ The results are shown in Fig.~\ref{perc-scal}
and recorded in Table~\ref{perc-params}.

The formula in Eq.~(\ref{droplet_distribution}) used in this analysis
is only one example of a more general form of the scaling assumption
\cite{stauffer-79,stauffer-text}
	\begin{equation}
	n_A = A^{-\tau} f \left( X \right)
	\label{gen-scaling}
	\end{equation}
with $X = A^{\sigma} {\varepsilon}^{\varphi} / T$ and where $f(X)$ is
some general scaling function.\ \ This scaling function should be
valid on both sides of the critical point.\ \ For small $X$ ($T
\approx T_c$ and small $A$) and $\varepsilon > 0$, $f(X)$ will vary as
$\exp(-X)$ with $\sigma = 1 / ( \beta \delta ) = 1 / ( \gamma + \beta)
= 0.64$ for $d=3$ Ising systems or $0.45$ for $d=3$ percolation
systems and $\varphi = 1$.\ \ For large $X$ ($T$ far from $T_c$ or
large $A$) and $\varepsilon > 0$, $f(X)$ will vary as $\exp(-X)$ with
$\sigma=2/3$ for all three dimensional systems and with $\varphi = 2 \nu$;
where $\nu = 0.63$ for $d=3$ Ising systems and $\nu = 0.88$ for $d=3$
percolation lattices.

The fitting procedure using Eq.~(\ref{droplet_distribution}) returned
a value of $\sigma = 0.44 \pm 0.01$ and $\tau = 2.192 \pm 0.003$ in
good agreement with other measurements, $\sigma = 0.45$ and $\tau =
2.18$ \cite{stauffer-text}.\ \ It is clear from these results that the
data examined here is in the small $X$, $\varepsilon > 0$ region where
the approximation of $f(X)$ given in Eq.~(\ref{droplet_distribution})
is valid.\ \ As with the EOS data, the errors quoted here are from the
fitting procedure.\ \ Systematic errors that arise from the use of
Fisher's scaling form and from the fitting regions in $A$ and $q$ are
on the order of $\sim10$\%.

The value of $c_0$ for the $L = 6$ lattice is in good agreement with
previous measures \cite{elliott-00.2}.\ \ The interpretation of the
change in $c_0$ with lattice size will be discussed below.

The values of $\Delta \mu$ for all lattices are close to zero, in
agreement with the fact that percolation calculations such as these
are at coexistence.

It is now a simple matter to follow the analysis described above using
Fisher's parameterization of the cluster distribution to determine
the ``phase diagrams'' for these percolation lattices.\ \ The
interpretation of these ``phase diagrams'' is not as simple.

        \begin{figure} [ht]
        \centerline{\psfig{file=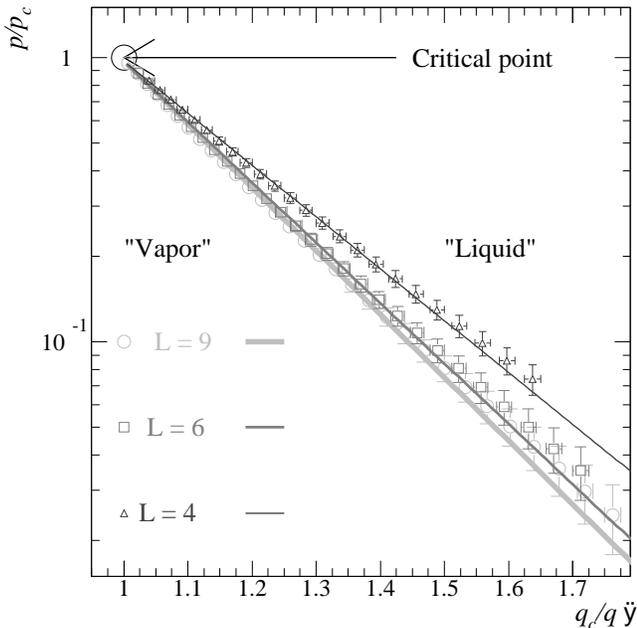,width=8.3cm,angle=0}}
        \caption{The reduced pressure-probability phase diagram: the
        points show calculations performed at the probabilities
        below the critical point and the lines show fits to the
        Clausius-Clapeyron equation.}
        \label{perc-pres-prob}
        \end{figure}

First the ``reduced pressure'' as a function of the inverse of the
``reduced probability'' $q/q_c$ is determined via
Eq.~(\ref{reduced_pressure}), and as usual for percolation studies $q$
replaces $T$ and $q_c$ replaces $T_c$.\ \ The results are shown in
Fig.~\ref{perc-pres-prob} where the points are fit with
Eq~(\ref{gugg-01}).\ \ This leads to an estimate of the ``enthalpy of
evaporation of a cluster'' given in Table~\ref{perc-thermo-table}.\ \
The values of $\Delta H$ are on the order of the values of $c_0$ and
increase with increasing $L$.

	\begin{table}
	\caption{``Thermodynamic'' properties of finite percolation lattices}
	\begin{tabular}{lllll}
	$L$ & $\Delta H$        & $\Delta E / A$ & $C^{F}_{c}$       & bonds$/$site \\
	\hline
	$9$ & $3.62 \pm 0.03$   & $2.7 \pm 0.1$  & $0.275 \pm 0.003$ & $2.67$       \\
	$6$ & $3.35 \pm 0.03$   & $2.2 \pm 0.1$  & $0.275 \pm 0.003$ & $2.50$       \\
	$4$ & $2.75 \pm 0.03$   & $1.8 \pm 0.1$  & $0.275 \pm 0.003$ & $2.25$       \\
	\end{tabular}
	\label{perc-thermo-table}
	\end{table}

        \begin{figure} [ht]
        \centerline{\psfig{file=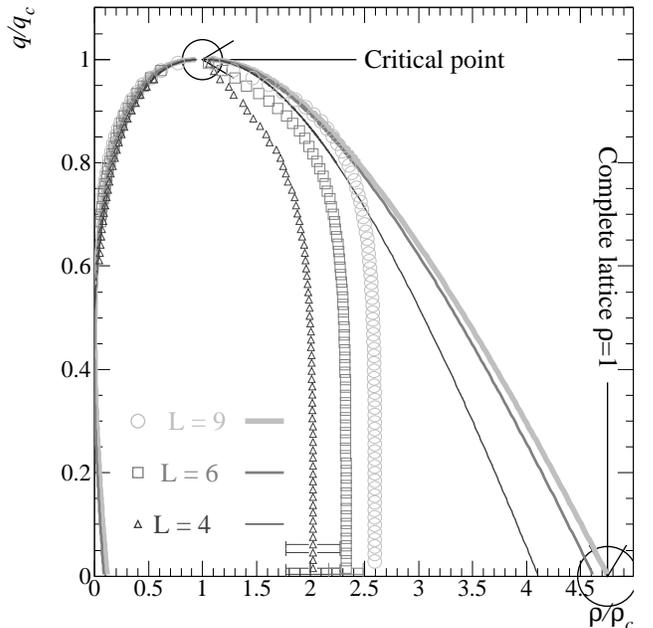,width=8.3cm,angle=0}}
        \caption{The points shown on the low density branch are from
        the calculations performed at the bond breaking probabilities
        below the critical point.\ \ The lines are a fit to and
        reflection of Guggenheim's equation.\ \ The points shown on
        the high density branch show the size of the largest fragment
        at a given value of $q$ normalized to the size of the largest
        fragment at $q_c$.}
        \label{perc-prob-dens}
        \end{figure}

To determine the ``energy of vaporization'' of a cluster $\Delta E$
the ideal gas approximation $pV = q$ is followed so that $\Delta E =
\Delta H - q$, where $T$ is replaced by $q$ in keeping with standard
practice in percolation work and $q$ is the average bond breaking
probability considered; $0.56 \pm 0.02$, $0.54 \pm 0.02$ and $0.53 \pm
0.02$ for $L = 9$, $6$ and $4$ respectively.\ \ The $\Delta E/A$ values
listed in Table~\ref{perc-thermo-table} were found by dividing $\Delta
H - q$ by the most probable cluster size ($1.15 \pm 0.05$, $1.25\pm
0.05$ and $1.25 \pm 0.05$ for $L = 9$, $6$ and $4$ respectively), this
puts $\Delta E$ on a ``per site'' basis.

The values of $\Delta E/A$ shown in Table~\ref{perc-thermo-table} are
nearly identical to the values of the surface energy coefficient
$c_0$, which is not surprising since for percolation on a simple cubic
lattice $c_0$ arises from the bonds broken to form the surface.\ \
Furthermore, the ``energy of vaporization'' is approximately equal to
the number of bonds per lattice site (also shown in
Table~\ref{perc-thermo-table}), a strong indication that the $\Delta
E/A$ calculated here is the ``bulk binding energy'' of the lattice in
question.\ \ The value of $\Delta E/A$ decreases with the size of the
lattice because the percolation calculations were performed for open
boundary conditions.

The compressibility factor at the critical point was determined via
Eq.~(\ref{comp-fac-eq}), the results are shown in
Table~\ref{perc-thermo-table}.\ \ From $C^{F}_{c}$, $q_c$ and
${\rho}_c$ (determined below) the ``pressure'' at the critical point
can be found.\ \ The resulting values of $p_c$ are shown in
Table~\ref{perc-critical-point}, but the interpretation of these
values is an open question.

Following the thermodynamic treatment of the percolation results, the
reduced probability versus ``reduced density'' phase diagram is
produced via Eq.~(\ref{reduced_density}).\ \ This leads to the points
shown in Fig.~\ref{perc-prob-dens}.\ \ These points are then fit to
Guggenheim's empirical formula, Eq.~(\ref{gugg-eq}), with $\beta =
0.43 \pm 0.01$ (in good agreement with text book values $0.41$
\cite{stauffer-text}) from Eq.~(\ref{beta}).\ \ These results are shown
for each lattice by the solid lines in Fig.~\ref{perc-prob-dens}.

While it is not clear what density this plot describes, some insight
can be gained by noting that the ``liquid'' branch reaches
${\rho}/{\rho}_c \sim 4$ to $\sim4.5$ at $q = 0$.\ \ Assuming that at
$q = 0$ $\rho = 1$, since no bonds are broken, then ${\rho}_c \sim
0.22$ to $\sim 0.25$, which is approximately the percentage of bonds
broken at the critical point.\ \ Thus it seems that the density in
Fig.~\ref{perc-prob-dens} is related to the number of broken bonds.\ \
It is also noted that for $q=0$ the vapor branch of the coexistence
curve shows ${\rho}/{\rho}_c > 0$, this serves as an illustration of
the magnitude of the error associated with this procedure.

It is also possible to directly explore the behavior of the reduced
density of the ``liquid,'' at least in the larger system.\ \ This is
done by normalizing the size of the largest cluster at a given value of
$q$ to the size of the largest cluster at the critical point
$A_{\text{max}}(q)/A_{\text{max}}(q_c)$.\ \
Figure~\ref{perc-prob-dens} shows that for the $L=9$ lattice, the
measured normalized density of the liquid tracks along the coexistence
curve predicted by Guggenheim's empirical formula and the reduced
density of the vapor.\ \ For $q/q_c < 0.75$ the effects of the finite
size of the lattice are observed and the measured reduced density of
the liquid deviates from the coexistence curve.\ \ The effects of
finite size are more evident in the smaller lattices where there is
little or no agreement between the measured reduced density of the
liquid and the coexistence curves.\ \ Effects of finite size on the
largest cluster such as these have been observed previously
\cite{elliott-94}.

        \begin{figure} [ht]
        \centerline{\psfig{file=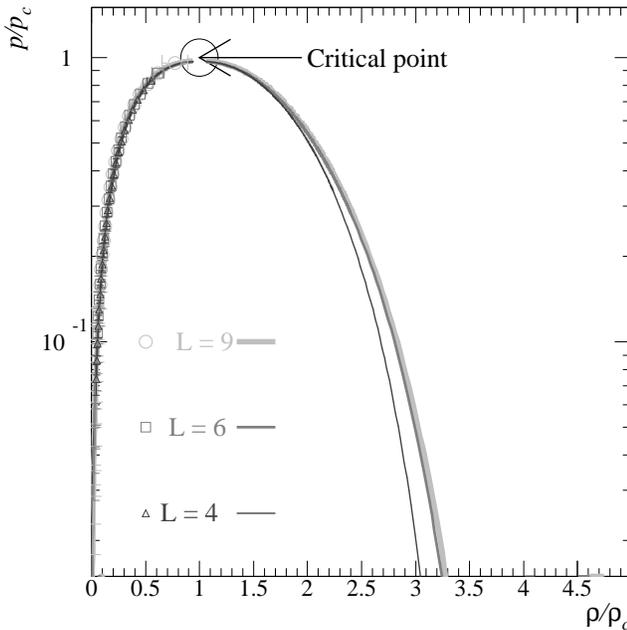,width=8.3cm,angle=0}}
        \caption{The points are calculations performed at the bond
        breaking probabilities below the critical point and the lines
        are the results of the fits from the previous sections.}
        \label{perc-pres-den}
        \end{figure}

For the sake of completeness, the ``reduced pressure'' versus
``reduced density'' projection of the phase diagram is shown in
Fig.~\ref{perc-pres-den}.

\end{document}